\documentclass[twocolumn,showpacs,preprintnumbers,amsmath,amssymb]{revtex4}
\usepackage{graphicx}
\usepackage{dcolumn}
\usepackage{bm}

\begin{document}

\title{The Vector Analyzing Power in Elastic Electron-Nucleus Scattering}
\author{E. D. Cooper}
\affiliation{University College of the Fraser Valley
33844 King Road, Abbotsford, BC, Canada  V2S 7M8} 
\author{C.J. Horowitz\email{horowit@indiana.edu}} 
\affiliation{Nuclear Theory Center and Department of Physics, 
             Indiana University,  Bloomington, IN 47405}

\date{\today}

\begin{abstract}
The vector analyzing power $A_n$ is calculated for elastic electron scattering from a variety of spin zero nuclei at energies from 14 MeV to 3 GeV.  Time reversal symmetry insures that $A_n$ vanish in first Born approximation.  Therefore $A_n$ depends on Coulomb distortions and can be large for scattering from heavy nuclei.  The vector analyzing power is a potential source of systematic error for parity violation experiments.  We find that $A_n=-0.361$ ppm for the kinematics of the Parity Radius Experiment (PREX) involving 850 MeV electrons scattering at six degrees from $^{208}$Pb.  This is comparable to the parity violating asymmetry.  However for HAPPEX He involving 3 GeV electrons scattering on $^{4}$He we find that $A_n$ is very small.   
\end{abstract}
\maketitle
%
\section{Introduction}
\label{intro}
Recent measurements of the vector-analyzing power $A_n$ in polarized electron-proton scattering probe nucleon structure and test theory \cite{sample,an1,rm}.  The analyzing power is a parity even, time reversal odd correlation between the electron spin and the target and electron momenta.  The time reversal symmetry of Quantum Electrodynamics insures that $A_n$ vanishes in first Born approximation.  Therefore, a nonzero measurement of $A_n$ is a direct observation of two (or more) photon exchange contributions.   These two-photon contributions are very interesting.  For example, the discrepancy in the ratio of the electric to magnetic form factors of the proton $G_E^p/G_M^p$, obtained via polarization transfer and Rosenbluth separation, may be due to the neglect of two-photon contributions \cite{blunden,guichon}.

The analyzing power for electron-proton scattering is of order the fine structure constant $\alpha$.  In contrast, electron scattering from a heavy nucleus has large Coulomb distortion effects and these give analyzing powers of order $Z\alpha$ where $Z$ is the nuclear charge.  These large analyzing powers are used to measure the polarization of low energy electrons \cite{mottpol}.  At higher energies they can be a source of systematic errors for parity violation experiments.  If a primarily longitudinally polarized electron beam has a small transverse polarization, this may couple with the analyzing power and generate a false asymmetry.  

The Parity Radius Experiment (PREX)\cite{prex, bigpaper} aims to measure the neutron radius of $^{208}$Pb accurately and model independently via parity violating electron scattering.  However, the analyzing power for scattering from Pb is relatively large because of the large charge.  This could produce a potentially serious systematic error.  Therefore, PREX may measure the analyzing power to control the error.  In this paper we calculate the expected $A_n$ for PREX kinematics.  In addition the HAPPEX He experiment is measuring parity violating electron scattering from $^4$He at 3 GeV in order to determine strange quark contributions to the electric form factor of the nucleon \cite{happexhe}.  This experiment will also measure $A_n$ from $^4$He.   

There have not been many calculations of $A_n$ at higher energies.  In this paper we accurately calculate analyzing powers for electron scattering from a variety of spin zero nuclei over a large range of laboratory energies from 14 MeV to 3 GeV.  We solve the Dirac equation to calculate the analyzing power to all orders in $Z\alpha$.  This turns out to be a substantial numerical challenge as discussed in Section \ref{formalism}.  In between photon exchanges, the nucleus can be in a variety of excited states.  However, only the ground state generates a coherent contribution proportional to $Z\alpha$.  The other excited states may only contribute at order $\alpha$.  Therefore the major approximation of this work is to assume that the nucleus remains in its ground state at all times.  This allows the scattering to be described with a simple Dirac equation.       

\section{Formalism}
\label{formalism}
In this section we describe the numerical procedure used in the code RUNT for calculating $A_n$.  First we discuss the treatment of recoil.  Next we discuss the numerical techniques necessary to calculate the scattering amplitude with enough accuracy to extract $A_n$.  Since $A_n$ is often small this is a very demanding numerical calculation.  

We wish to describe electron scattering from a spin zero nucleus.   We include recoil or center of mass corrections using the procedure of ref. \cite{ByronTim_smooth_prop}.  This approach ends up with a Dirac equation in the relative co-ordinate, and an effective kernel which is a recoil factor (similar to a reduced mass in the non-relativistic picture) times the potential. Corrections can be made to this picture, from higher order effects, but they are not included here.  With this approach, and our approximation that the nucleus remains in its ground state, the scattering can be described by the Dirac equation including the Coulomb potential $V_c(r)$ from the ground state charge distribution of the target. 

The time-independent Dirac equation is,
\begin{equation}
[- i \alpha \cdot \nabla + \beta m_e + \frac{E_t}{E_t+E_e} V_c(r)\ -\ E_e ] \Psi( r) = 0,
\label{dirac}
\end{equation}
where $E_t$ ($E_e$) is the total energy (kinetic + rest mass) of the target (electron) in the centre of momentum frame and $m_e$ is the electron mass.  

Since for large values of $r$, $V_c(r)$ goes like $1/r$ our phase shifts come from matching our radially integrated wavefunctions to regular and irregular Dirac-Coulomb functions in the standard way.

To calculate the scattering observables from the phase shifts there are two scattering amplitudes,
\begin{equation}		 
 f(\theta) = \sum_{l=0}^\infty  A_l P_l(\cos\theta),                            
\label{fsum}
\end{equation}
\begin{equation}
 g(\theta) =\sum_{l=1}^\infty	  B_l P_l^1 (\cos\theta),                  
\label{gsum}
\end{equation}
where,
\begin{equation}
A_l = (l+1) \eta_l^+ + l \eta_l^-,
\label{A2}          
\end{equation}
and
\begin{equation}
B_l = \eta_l^- - \eta_l^+,
\label{A3}
\end{equation}

\begin{equation}
\eta_l^\pm = \frac{\exp[2i\delta(l, j=l\pm 1/2)] -1}{2ik},        
\end{equation}
and $k$ is the electron's momentum in the centre of momentum frame.

The phase shift $\delta(l,j)$ contains a part that would be obtained from scattering from a pure Coulomb potential(calculated analytically), plus a piece that is obtained numerically from matching to Dirac Coulomb wavefunctions outside the charge distribution (typically from 10 to 18 fm from the centre of the nucleus), and has the property that it goes rapidly to zero with increasing $l$, whereas the pure Coulomb phase shift goes to zero so slowly that the series in the expressions for $f(\theta)$ and $g(\theta)$ diverge and converge painfully slowly respectively.  The standard approach to summing this series is well known, being first presented in ref\cite{yrw}.   

From the amplitudes we can calculate the cross section and analyzing power,
\begin{equation}
d\sigma /d\Omega = |f(\theta)|^2  + |g(\theta)|^2
\end{equation}
and
\begin{equation}
A_n = \frac{2{\rm Im} [ f^*(\theta) g(\theta) ]}{|f(\theta) |^2  + |g(\theta) |^2} .     
\label{a8}
\end{equation}
The analyzing power gives the fractional difference in cross section for scattering to the left versus right for electrons that are initially polarized transverse to the scattering plane.  

The problem of calculating the analyzing power for electron scattering by numerically solving the Dirac equation is not an easy one. The analyzing power is very small for light nuclei (of the order $10^{-11}$), so calculating it involves calculating $f(\theta)$ and $g(\theta)$ to a very high precision. For heavier nuclei the cross section drops rapidly, so calculating an analyzing power from amplitudes that have already dropped several orders of magnitude is almost as difficult.

In this paper we have overcome the problem using several techniques which are modifications to the standard approach.  We present these techniques to help anyone trying to reproduce our results. 

\subsection{Radial Integration Modifications}

The code RUNT is used for most of the calculations \cite{RUNT}. The results are checked with an independent code ELASTIC \cite{ELASTIC, ELASTIC2}.  RUNT uses the Numerov algorithm.  At each radial step of length $h$ this introduces an error that goes like $h^6$, and after integrating over a certain distance, the accumulated error goes like $h^4$.  In practice, using a small enough step size to get the required precision involved very large arrays, and we also found that round-off (or truncation) error of the double precision numbers prevented us achieving the precision we needed.   

To use the Numerov algorithm we convert the coupled first order Dirac equations into a single second order differential equation, 
\begin{equation}
{y^{\prime\prime}(x)} = f(x) y(x)
\end{equation}
where
\begin{equation}
f(x) = -k^2 + U(x) + \frac{l(l+1)}{x^2},
\end{equation}
and the Schrodinger-equivalent potential $U(x)$ is,
\begin{equation}
U(x)=-\frac{\kappa}{r}(\frac{U^\prime_c(x)}{E_e+m_e - U_c(x)}) + 2E_e U_c(x) -U_c(x)^2 
\nonumber
\end{equation}
\begin{equation}
\ \ \ \ 
 + \frac{3}{4}(\frac{U^\prime_c(x)}{E_e+m_e-U_c(x)})^2 + \frac{U^{\prime\prime}_c(x)}{2(E_e+m_e-U_c(x))}.
\label{scheqiv}
\end{equation}
Here $U_c(x)=E_t V_c(x)/(E_t+E_e)$, $U^\prime_c(x)=dU_c(x)/dx$, and $\kappa$ is the usual Dirac quantum number.

The Numerov method is a three point formula:
\begin{equation}
w(x+h) = \frac{2 + 10T(x)}{1-T(x)} w(x) - w(x-h).
\label{eq3}
\end{equation}
where $T(x)=h^2 f(x)/12$ and,  
\begin{equation}
w(x) = (1-T(x))y(x).
\end{equation}
It is self-starting for $l\neq 1$ and requires no power series for the first step. However for p-waves $l=1$ it does not start by itself.  We tried several ways of starting the integration but they all ended up resulting in p-wave phase shifts less accurate than the other phase shifts. Empirically we found the best way, which was fortunately good enough, to be as follows.

We make use of the fact that near the origin, any physical potential will be an even function of $r$ and therefore can be approximated as constant. When $U(x)$ is constant the solution is, 
\begin{equation}
y(x) = A j_1(k^\prime x)
\end{equation}
Where 
\begin{equation}
{k^\prime}^2 = k^2 - U(h).  
\end{equation}
Thus for the first step we take 
\begin{equation}
y(2h) = y(h) j_1(2k^\prime h)/j_1(k^\prime h)
\end{equation}
where $j_1(x)$ is a regular spherical Bessel function. Thus the first step is done this way, and then we use 
use Eq. \ref{eq3} to integrate out.  This helped, but still did not result in a method that was accurate enough, however.

The next step was to recognize that the kinetic energy of the electrons is much higher than the potential they feel. Thus we are calculating barely perturbed plane waves. Under such circumstances we used two of the techniques discussed in  ref\cite{thorlacius}.

The first technique is to modify Eq. \ref{eq3} so that it becomes exact in the limit of constant f(x).  This results in the formula,
\begin{equation}
w(x+h) = 2\cosh( \sqrt{12T} ) w(x) - w(x-h)   
\label{eq10}
\end{equation}
This formula still has an error that goes like $h^6$, but it is smaller than in Eq. \ref{eq3}, and this allowed us to gain at least two more significant figures, but this still wasn't accurate enough.

The second technique is to look at where Eqs. \ref{eq3} and/or \ref{eq10} come from, and see that the errors involved in using these formulae to integrate over a fixed range (typically $0\rightarrow 18$ fm in our case), go like $c_1 h^4 + c_2 h^6 + c_3 h^8 +...$. Under such circumstances Richardsonian extrapolation works very well.  We found in practice for this problem that two iterations of the Richardsonian method were optimal (two iterations gained us the accuracy we needed, and more than this did not help as we were then at the round-off error limit).
 
With these techniques, we found that our phase shifts could be calculated to the desired accuracy by matching to Dirac Coulomb wave functions calculated by the method of Barnet and Steed \cite{comfort}.

\subsection{Phase-Shift Summation} 

The standard approach in electron scattering, due to Yennie, Ravenhall and Wilson \cite{yrw} is to relate the (divergent) sum over Legendre Polynomials in equations \ref{fsum} and \ref{gsum} for the scattering amplitudes $f(\theta)$, $g(\theta)$  with a sum over Legendre polynomials for the functions $(1-\cos(\theta))^n  f(\theta)$ and $(1-\cos(\theta))^n  g(\theta)$. These new sums converge more quickly, but to extract the original amplitudes we need to divide by $(1 -\cos(\theta))^n$ which blows up near $\theta=0$. Thus we have to do our sums very very accurately at forward angles. $n$ is some small integer, usually taken as 2,3,4 or 5.  This method works very well away from the forward angles, but is problematic there.

Since we want to calculate accurately near forward angles, we use a different mathematical device, inspired by the physics of screening. In theory electron scattering from a point Coulomb potential gives a divergent partial wave series, but in practice the screening of the electron clouds will (for a neutral atom target), cause the partial wave series to converge.

We crudely model this by multiplying the partial wave amplitudes $A_l$ and $B_l$ in equations \ref{A2}, \ref{A3}, by a convergence factor $\exp(-(l/a)^2)$ where $a$ is taken to be large enough that our answer does not change significantly if $a$ is increased further.  

In practice we summed these term up to about $l=6a$ (by which time the convergence factor has become very small) , and used a value of $a$ somewhere between  500 to 500,000 (depending on electron energy and target charge).  We do caution that this method is NOT as numerically accurate at large angles as the method of Yennie, Ravenhall and Wilson. The two methods complement each other, and there is a large overlap at middle angles where they give the same result.

\subsection{
Tests of Calculation}

We list some of the tests of our results.  First, our results were obtained using two independent codes RUNT and ELASTIC.  Second we reproduce the analyzing powers and cross sections for scattering with 14 MeV electrons presented in ref. \cite{14MeV}.  Finally, for light (enough) electron mass $m_e$, $A_n$ is proportional to $m_e$.  Should the electon mass get 10x heavier, and the centre-of-mass wave-vector remain the same, then the analyzing power should get 10X bigger.  We explored this scaling by multiplying the electron rest-mass by 10, 100 and 1000 and re-adjusted the incident electron kinetic energy so as to obtain the same centre of mass wave vector as was obtained by having the correct electron mass.  We tested this for Lead at 200 MeV and for He at 3GeV (ignoring recoil effects) and found that provided the incident kinetic energy was not adjusted by more than 20\% the scaling of the analyzing power worked to better then 1 \%.  This test is especially re-assuring since numerically calculating analyzing powers using 100x heavier electrons is much easier.  All calculations presented here do use the physical electron mass, but others trying to reproduce the numbers may find it easier to use a heavier electron mass.

\section{Results}
\label{results}

In this section we present results for $A_n$ for scattering from a variety of spin zero nuclei at energies from 200 MeV to 3 GeV.  The experiment SAMPLE has measured $A_n$ for electron-proton scattering at 200 MeV and scattering angles from 130 to 170 degrees\cite{sample}.  It is interesting to compare this e-p result to scattering from $^{208}$Pb.  In Fig. 1 we show our $e-^{208}$Pb cross section.  We use the experimental charge density of ref. \cite{sog}.  Figure 2 shows our prediction for $A_n$.  For example, at a scattering angle of 146 degrees we find $A_n=-890$ ppm.  This is very large compared to the SAMPLE result of $-15.4\pm 5.4$ ppm for scattering from the proton.  Our result for Pb is much larger because of the large charge of Pb which greatly increases Coulomb distortions.  Also the proton $A_n$ is somewhat reduced by a magnetic moment contribution \cite{rm} that is absent for scattering from a spin zero nucleus.  Figure 3 shows $A_n$ at angles near 180 degrees where $A_n$ is very large $\approx 0.2$.  However, the elastic cross section is very small.  Note that at lower energies, Mott polarimeters use this back angle peak in $A_n$.

\vspace{0.02in}
\begin{figure}[ht]
\begin{center}
\includegraphics[width=2.5in,angle=270,clip=false]{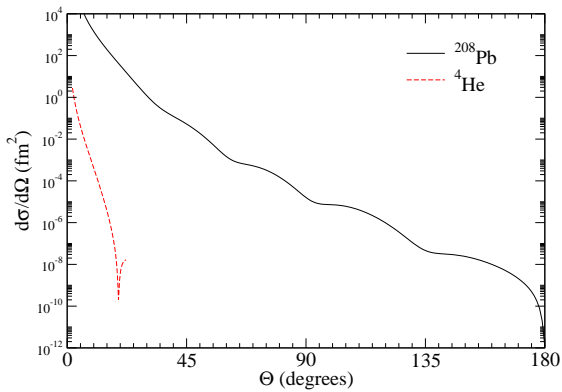}
\caption{(Color online) Differential cross section for elastic electron scattering versus scattering angle.  The solid line is for $^{208}$Pb at 200 MeV while the dashed line is for $^4$He at 3 GeV.}
\label{Fig1}
\end{center}
\end{figure}

\vspace{0.02in}
\begin{figure}[ht]
\begin{center}
\includegraphics[width=2.5in,angle=270,clip=false]{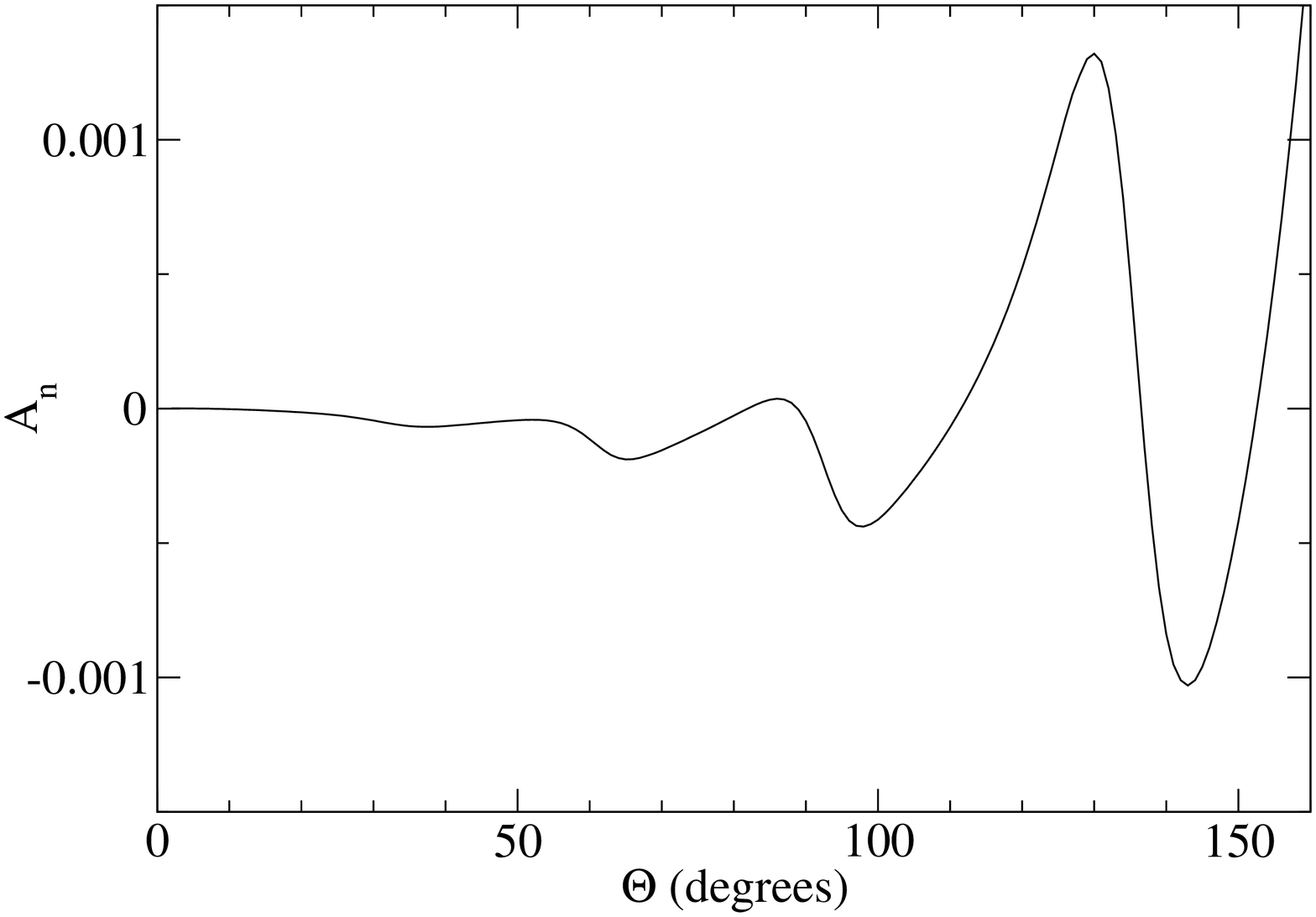}
\caption{The vector analyzing power $A_n$ for elastic electron scattering from $^{208}$Pb versus scattering angle at an energy of 200 MeV. }
\label{Fig2}
\end{center}
\end{figure}

\vspace{0.02in}
\begin{figure}[ht]
\begin{center}
\includegraphics[width=2.5in,angle=270,clip=false]{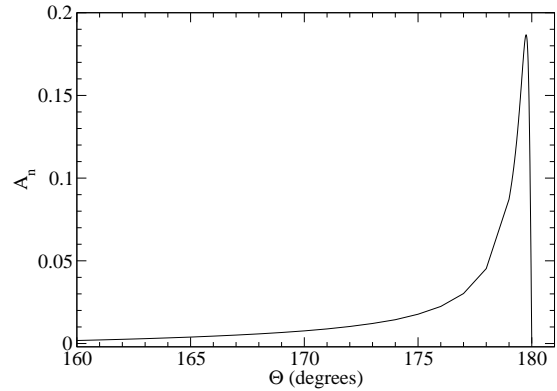}
\caption{The vector analyzing power $A_n$ as in Fig. 2 except at very large scattering angles.}
\label{Fig3}
\end{center}
\end{figure}

The Parity Radius Experiment (PREX) should measure the parity violating asymmetry $A$ for elastic electron scattering from $^{208}$Pb at 850 MeV and scattering angles near six degrees \cite{prex}.  The goal of PREX is to determine the neutron radius to 1 \% \cite{bigpaper}.  The weak charge of a neutron is much larger than that of a proton.  Therefore $A$ is very sensitive to the neutron density.  However, Coulomb distortions reduce $A$ by approximately 30 \% and are crucial for the interpretation of the experiment \cite{ELASTIC}.  A measurement of the vector analyzing power $A_n$ directly probes these distortions since $A_n$ is driven by Coulomb distortions and it vanishes in first Born approximation.  Figure 4 shows our predictions for $A_n$ at 850 MeV.  At a scattering angle of six degrees $A_n=-0.361$ ppm, see table I.  This is comparable to $A$ and thus a potential source of systematic error for the parity measurement.  However, this error can be controlled by a measurement of $A_n$ during the PREX experiment.      
 
\vspace{0.02in}
\begin{figure}[ht]
\begin{center}
\includegraphics[width=2.5in,angle=270,clip=false]{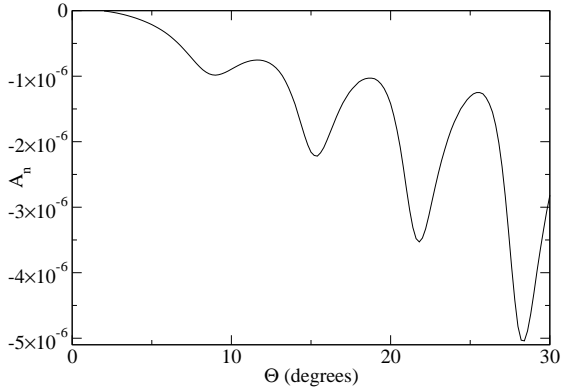}
\caption{The vector analyzing power $A_n$ for elastic electron scattering from $^{208}$Pb versus scattering angle at an energy of 850 MeV. }
\label{Fig4}
\end{center}
\end{figure}

\begin{table}
\caption{Differential cross section and analyzing power $A_n$ for elastic electron scattering from $^{208}$Pb at 850 MeV.}
\begin{tabular}{lcc}
$\theta$ (deg.)& $d\sigma/d\Omega$ (fm$^2$) & $A_n$ \\
4 &	1.76E+03 &	-1.14E-07\\
4.2 &	1.33E+03 &	-1.31E-07\\
4.4 &	1.00E+03 &	-1.49E-07\\
4.6&	7.59E+02 &	-1.69E-07\\
4.8&	5.76E+02 &	-1.90E-07\\
5&	4.37E+02 &	-2.13E-07\\
5.2&	3.33E+02 &	-2.38E-07\\
5.4&	2.53E+02 &	-2.65E-07\\
5.6&	1.93E+02 &	-2.94E-07\\
5.8&	1.47E+02 &	-3.26E-07\\
6&	1.12E+02 &	-3.61E-07\\
6.2&	8.50E+01 &	-3.99E-07\\
6.4&	6.48E+01 &	-4.41E-07\\
6.6&	4.94E+01 &	-4.85E-07\\
6.8&	3.77E+01 &	-5.33E-07\\
7&	2.89E+01 &	-5.85E-07\\
7.2 &	2.22E+01 &	-6.39E-07\\
7.4&	1.71E+01 &	-6.94E-07\\
7.6&	1.33E+01 &	-7.50E-07\\
7.8&	1.04E+01 &	-8.05E-07\\
8&	8.17E+00 &	-8.56E-07\\
8.2&	6.51E+00 &	-9.01E-07\\
8.4&	5.25E+00 &	-9.37E-07\\
8.6&	4.29E+00 &	-9.64E-07\\
8.8&	3.55E+00 &	-9.79E-07\\
9&	2.97E+00 &	-9.84E-07\\
9.2&	2.51E+00 &	-9.78E-07\\
9.4 &	2.15E+00 &	-9.64E-07\\
9.6&	1.85E+00 &	-9.44E-07\\
9.8&	1.60E+00 &	-9.20E-07\\
10&	1.39E+00 &	-8.93E-07\\
\label{tableone}
\end{tabular}

\end{table}

Figure 5 illustrates how $A_n$ depends on nuclear charge $Z$.  It shows results for $^{16}$O, $^{40}$Ca, $^{90}$Zr and $^{208}$Pb at 850 MeV.  At forward angles the magnitude of $A_n$ increases rapidly with increasing target $Z$.  For $^{16}$O, $A_n$ is small except in the diffraction minima where the denominator in Eq. \ref{a8} is small.  For light nuclei the diffraction minima are sharp and deep.  They become broader and less deep in heavy nuclei, see Fig. 1.  This is reflected in the structure observed for $A_n$.
\vspace{0.02in}
\begin{figure}[ht]
\begin{center}
\includegraphics[width=2.5in,angle=270,clip=false]{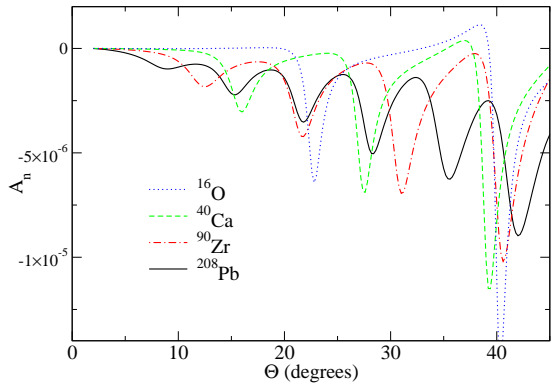}
\caption{(Color online) The vector analyzing power $A_n$ for elastic electron scattering from $^{16}$O (blue dotted line), $^{40}$Ca (green dashed line), $^{90}$Zr (red dot-dashed line), and $^{208}$Pb (black solid line) versus scattering angle at an energy of 850 MeV. }
\label{Fig5}
\end{center}
\end{figure}

Finally, HAPPEX He is measuring the parity violating asymmetry $A$ for elastic electron scattering from $^4$He at 3 GeV and angles from six to nine degrees.  To control systematic errors, $A_n$ will also be measured during HAPPEX He.  Figure 6 shows our predictions for $^4$He at 3 GeV using the three parameter Fermi charge density from ref. \cite{3pf}.  We find $A_n$ to be very small, of order $10^{-10}$ and it passes through zero near 7.7 degrees.  These results are so small compared to $A$ because of the small $Z$ of $^4$He, the high energy, and cancellations near the zero crossing.  For completeness, we have collected in Tables I and II our $A_n$ predictions for the kinematics of the HAPPEX He and PREX experiments.  
\vspace{0.02in}
\begin{figure}[ht]
\begin{center}
\includegraphics[width=2.5in,angle=270,clip=false]{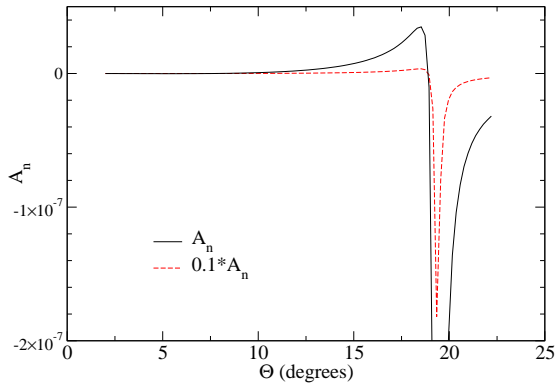}
\caption{(Color online) The vector analyzing power $A_n$ for elastic electron scattering from $^{4}$He (solid line) versus scattering angle at an energy of 3 GeV. Also shown is $A_n$ divided by 10 (red dashed line) to illustrate structure in the diffraction minimum.}
\label{Fig6}
\end{center}
\end{figure}
\begin{table}
\caption{Differential cross section and analyzing power $A_n$ for elastic electron scattering from $^{4}$He at 3 GeV.}
\begin{tabular}{lcc}
$\theta$ (deg.)& $d\sigma/d\Omega$ (fm$^2$) & $A_n$ \\
5.88&	1.72E-02&	-1.29E-10\\
6.08&	1.41E-02&	-1.25E-10\\
6.28&	1.16E-02&	-1.18E-10\\
6.49&	9.57E-03&	-1.09E-10\\
6.69&	7.91E-03&	-9.81E-11\\
6.9	&6.55E-03&	-8.41E-11\\
7.1&	5.43E-03&	-6.72E-11\\
7.31&	4.51E-03&	-4.73E-11\\
7.51&	3.75E-03&	-2.40E-11\\
7.71&	3.12E-03&	2.84E-12\\
7.92&	2.60E-03&	3.35E-11\\
8.12&	2.17E-03&	6.82E-11\\
8.33&	1.81E-03&	1.07E-10\\
8.53&	1.51E-03&	1.51E-10\\
8.73&	1.26E-03&	2.00E-10\\
8.94&	1.05E-03&	2.54E-10\\
9.14&	8.80E-04&	3.13E-10\\
\label{tabletwo}
\end{tabular}

\end{table}

\section{Conclusions}
\label{conclusions}
The analyzing power $A_n$ gives the fractional difference in cross section for scattering to the left versus right for electrons that are initially polarized transverse to the scattering plane.  It is a potential source of systematic error for parity experiments.  Because of time reversal symmetry, $A_n$ vanishes in first Born approximation and is driven by Coulomb distortions.  We have calculated $A_n$, for scattering from a variety of spin zero nuclei, by accurately solving the Dirac equation.  This is a numerically demanding task.  Very roughly away from diffraction minima, $A_n$ scales with $Z\alpha m_e/E_e$ where $Z$ is the target charge, $\alpha$ the fine structure constant, $m_e$ the electron mass and $E_e$ the electron energy.  Therefore $A_n$ can be large for heavy nuclei.  For the kinematics of the Parity Radius Experiment (PREX), 850 MeV electron scattering from $^{208}$Pb at six degrees, we find $A_n=-0.361$ ppm which is comparable to the parity violating asymmetry.  However for the kinematics of the HAPPEX He experiment, 3 GeV scattering from $^{4}$He, we find $A_n$ to be very small, of order $10^{-10}$.

This work was supported in part by DOE grant:  DE-FG02-87ER40365.

\end{document}